\documentclass[preprint,10pt]{elsarticle}

\makeatletter
\def\ps@pprintTitle{%
\let\@oddhead\@empty
\let\@evenhead\@empty
\def\@oddfoot{\centerline{\thepage}}%
\let\@evenfoot\@oddfoot}
\makeatother

\journal{Physica A}

\usepackage{graphicx,bbm,setspace,amssymb,amsfonts,amsmath,mathtools,mathrsfs,stmaryrd,amsthm,bm,etoolbox,comment,hyperref,url,caption,subcaption,color,enumitem,algorithm,booktabs,microtype,nicefrac,lingmacros,nccmath,tree-dvips,tikz-cd}
\usepackage[noend]{algpseudocode}
\usepackage{algorithm}

\usepackage[colorinlistoftodos]{todonotes}
\usepackage[utf8]{inputenc} 
\usepackage[T1]{fontenc}    

\usepackage[noend]{algpseudocode}

\newtheorem{thm-defn}[theorem]{Theorem/Definition}

\theoremstyle{definition}

\theoremstyle{remark}

\newcommand{\ignore}[1]{}{}

\begin{document}

\begin{frontmatter}

\title{On financial market correlation structures and diversification benefits across and within equity sectors}
   
\author[label1]{Nick James} \ead{nick.james@unimelb.edu.au}
\author[label2]{Max Menzies} \ead{max.menzies@alumni.harvard.edu}
\author[label3]{Georg A. Gottwald} \ead{georg.gottwald@sydney.edu.au}
\address[label1]{School of Mathematics and Statistics, University of Melbourne, Victoria, Australia}
\address[label2]{Beijing Institute of Mathematical Sciences and Applications, Tsinghua University, Beijing, China}
\address[label3]{School of Mathematics and Statistics, University of Sydney, NSW, Australia}

\begin{abstract}
We study how to assess the potential benefit of diversifying an equity portfolio by investing within and across equity sectors. We analyse 20 years of US stock price data, which includes the global financial crisis (GFC) and the COVID-19 market crash, as well as periods of financial stability, to determine the `all weather' nature of equity portfolios. We establish that one may use the leading eigenvalue of the cross-correlation matrix of log returns as well as graph-theoretic diagnostics such as modularity to quantify the collective behaviour of the market or a subset of it. We confirm that financial crises are characterised by a high degree of collective behaviour of equities, whereas periods of financial stability exhibit less collective behaviour. We argue that during times of increased collective behaviour, risk reduction via sector-based portfolio diversification is ineffective. Using the degree of collectivity as a proxy for the benefit of diversification, we perform an extensive sampling of equity portfolios to confirm the old financial adage that 30-40 stocks provide sufficient diversification. Using hierarchical clustering, we discover a `best value' equity portfolio for diversification consisting of 36 equities sampled uniformly from 9 sectors. We further show that it is typically more beneficial to diversify across sectors rather than within. Our findings have implications for cost-conscious retail investors seeking broad diversification across equity markets.

\end{abstract}


\end{frontmatter}


\section{Introduction}
\label{Introduction}
Financial market structure and behaviour are notoriously difficult to describe and predict. Over the last 100 years, countless mathematical models and intuitive rules have been developed to predict the behaviour of individual assets as well as broader market trends. In 1952, Markowitz \cite{Markowitz1952} revolutionised the study of financial markets and the practice of asset selection by arguing that diversification across many assets provides superior risk reduction to the optimal selection of individual assets. The idea of diversification relies on disentangling the risk of a particular financial asset into the risk of the market, the so called systematic risk, which an investor cannot control, and the individual risk of an asset, the so called unsystematic risk, which is assumed to be uncorrelated to the systematic risk of the market. Diversification amounts to averaging out the unsystematic risk by investing in a sufficient number of individual assets, leaving an investor exposed to only the inherent systematic risk of the market. The benefit of diversification is intimately tied to the notion that the price of an asset can be decomposed into a (noisy) collective market component and an idiosyncratic noisy component which is uncorrelated to the collective behaviour of the market \cite{EDERINGTON1993, Balduzzi2001, Andersen2007}. By analysing data from a 20-year period of 339 US equities, we aim to shed some light on how well this separation of the risk into a collective market component and into an individual component holds across time, and how diversification benefits vary when investing in different sub-collections of the market. We pay particular attention to the traditional method of diversifying across industry sectors, and study how beneficial this approach is in diversifying an equity portfolio. Until the last several decades, active investment management has been dominated by fundamental investors who make investment decisions based on the future earning potential of companies, relative to their current valuations. The correlation between the prosperity of companies in different sectors and that of the overall economy varies significantly. For example, equities classified in the Information Technology, Financials, Energy and Materials sectors often thrive during periods of economic growth. By contrast, sectors with more defensive earning profiles such as Healthcare, Utilities and Consumer staples tend to outperform during recessionary periods. Therefore, it is reasonable to expect that is more beneficial to diversify across sectors rather than within. However, this intuitive reasoning requires a thorough investigation backed up by data.

Ever since Markowitz' work, cross-correlation matrices of asset prices have been the key object of study in capturing market structure and the interdependencies of assets in the market or within a subset of the market such as equity sectors. These matrices' spectral properties encode important information about the overall market structure. To study evolutionary correlation structures, principal component analysis of the cross-correlation matrix was employed. In particular, the leading eigenvectors and eigenvalues were used to characterise the collective behaviour of the market. It was shown that a few components describe most of the observed variability of the market \cite{Pan2007,Wilcox2007,Fenn2011,Mnnix2012,Heckens2020}. Using random matrix theory, differences between cross-correlation matrices of stock price changes and random matrices can be used to uncover non-random aspects of the market \cite{Laloux1999,Plerou2002,Gopikrishnan2001,Fenn2011}. Network analysis, in which the stock market is viewed as a complex network where the cross-correlation matrix describes the coupling strength between individual assets, was used to find correlated groups of assets within the market \cite{Bonanno2003,Onnela2003,Onnela2004,Utsugi2004,Kim2005,Fiedor2014,Fiedor2014_2}. Correlation structures have often been studied using a variety of statistical and mathematical techniques to uncover various insights related to the evolution of global stock markets over time \cite{Song2011, Maslov2001, Noh2000}, and identify non-trivial temporal dependence structures \cite{Drod2000}. More generally, the econophysics community have used insights generated from evolutionary correlation structures to gain insights into a variety of arenas in the financial markets including equities \cite{James2022_inflation,james2021_MJW}, fixed income \cite{Driessen2003}, foreign exchange \cite{Ausloos2000,arjun} and cryptocurrencies \cite{James2021_crypto,JamescryptoEq, Wtorek2020, Drod2018,james2021_crypto2,Drod2019, Drod2020, Drod2020_entropy, James2022_physica_d, Chu2015,Sigaki2019}. The study of time-evolutionary correlation structures has also been of great use in other fields \cite{James2021_geodesicWasserstein,james2021_TVO,james2021_CovidIndia,james2021_hydrogen,james2021_olympics}. However, one must note that given the nonstationary and volatile nature of financial securities, they often exhibit heavy-tailed distributions \cite{Cerqueti2020, Wan2017, Stehlk2017, ChiaShangJamesChu1996, Chen2018, Cerqueti2019} and this can lead to limitations in the naive application of the Pearson correlation metric.  

The aim we set ourselves here is to find and employ a quantitative measure informing investors if diversifying their portfolio by investing in a larger number of equities will be beneficial for their risk reduction. The quantification of risk reduction is a difficult task, and highly definitional. We argue that diversification is beneficial if the unsystematic risk is sufficiently large compared to the systematic risk of the collective market. Hence diversification is beneficial in a market with a sufficiently low degree of collective behaviour, in which individual assets display a certain degree of independence. On the contrary, in a market which exhibits a high degree of collective behaviour, diversification may not lead to a significant reduction in the overall risk of the portfolio. Here we apply several complementary diagnostics to uncover dominant collective behaviour (or the lack thereof) of the market as a whole and in terms of individual sectors. We will use the leading eigenvalue of the cross-correlation matrix as a proxy for the collective behaviour of the market (or subset of the market) \cite{Avellaneda2019,Drod2020,Fenn2011}, with a larger value of it being indicative of stronger collective behaviour.  We further employ modularity, a diagnostic borrowed from complex network analysis, to probe into how far sectors function as mutually independent sets of equities. We find that all our metrics show the same signature: in times of crisis, such as the global financial crisis (GFC) in 2008/2009 or the 2020 market crash related to the COVID-19 pandemic, the market exhibits increased collective behaviour in which assets collectively react to overwhelming market and equity-specific unsystematic risk is swamped by the systematic risk of the market. By contrast, periods of sustained equity price growth (often referred to as bull market periods) are characterised by a lesser degree of collectivity, allowing for more efficient equity portfolio diversification.

There is a commonly held principle among investors that in order to diversify, a sufficient and perhaps optimal number of equities to hold in a portfolio is between 30 and 40 \cite{Fisher1970}. Many fund managers and individual investors may wish to limit their total number of held equities, either due to mandated restrictions in their investment policy statement \cite{Russellpolicy}, transaction fees, or complexity considerations of large portfolios. Hence finding the smallest number of equities which still allows for sufficient diversification is of paramount interest to investors. Using an exhaustive sampling strategy we aim to find evidence in the data of the 30-40 stock number rule, and how this rule is affected by the presence of sectors. Motivated by the results on the degree of collectivity, we measure the propensity of the market to allow for diversification by the reduction in the magnitude of the leading eigenvalue of the cross-correlation matrix associated with the respective portfolios. Perhaps unsurprisingly, we show that the precise selection of equities within a sector is less important than selecting a sufficient number of sectors to choose from. However, we will show that the data suggests that during the 20 year long period from 2001 to 2020 the anecdotal 30-40 equity rule does apply when investing across sectors. Interestingly, we show that a portfolio consisting of 36 equities sampled uniformly from 9 sectors provides comparable risk mitigation to a 90 equity portfolio, sampled uniformly from 10 sectors. Moreover, we show that risk reduction is less sensitive to the precise selection of equities within a sector once sectors are chosen. This supports the rationale behind recent trends in finance where diversification is promoted by investing in thematic areas. 

The paper is structured as follows. Section~\ref{sec:data} describes the US equity data used for our analysis. In Section~\ref{sec:marketstructure}, we study the market structure across and within sectors. We begin with a study of the leading eigenvalue of the correlation matrix to explore the collective behaviour of the whole portfolio comprised of all equities as well as within each equity sector. Periods of financial crises and of bull markets are clearly identified as increased and decreased collective behaviour, respectively. Periods of financial crisis are further characterised by an increase in the market's homogeneity. We augment the analysis by graph-theoretic-informed diagnostics and show that modularity can be used as another proxy for the degree of collectivity of the market, exhibiting the same temporal signatures as the leading eigenvalue of the cross-correlation matrix. In Section \ref{sec:portfoliosampling}, we turn to the more practical problem of studying the diversification benefit provided by diversifying across sectors. We verify that appropriately chosen combinations of 30-40 stocks across diverse sectors provides essentially as much diversification benefit as the entire market. 


\section{Data}
\label{sec:data}
We consider the daily stock prices of $N=339$ US equities from January 1 2000 to October 8 2020, i.e. a total of $T=5420$ data points. The data were downloaded from Bloomberg. The data periods include periods of major economic disruption such as the dot-com bubble in 2000/2001, the global financial crisis (GFC) in 2008/2009 with its subsequent severe market responses in 2010 and 2011, and the COVID-19 market crash in 2020, as well as more stable periods of equity market performance such as the sustained bull market from 2016-2019. The collection of equities can be divided into $M=11$ sectors according to the Global Industry Classification Standard (GICS). Each sector contains a different number of equities. The sectors are Communication Services (10 equities), Consumer discretionary (39 equities), Consumer staples (25 equities), Energy (18 equities), Financials (46 equities), Healthcare (44 equities), Industrials (55 equities), Information technology (36 equities), Materials (19 equities), Real estate (24 equities) and Utilities (23 equities). A list of the equities considered is given for completeness in \ref{sec:equity_securities}.


\section{The collective behaviour of the equity market}
\label{sec:marketstructure}

We will establish that one can measure the degree of collectivity of the market by certain spectral and graph-theoretic properties of the cross-correlation matrix of the log returns of the stock price data. These measures will be used in Section~\ref{sec:portfoliosampling} as a proxy for the benefit of diversification of a portfolio. 

We denote by $c_i(t)$, $i=1,\dots,N$, $t=0,\dots,T$ the multivariate time series of daily closing prices among our collection of $N$ equities. The multivariate time series of log returns, $r_i(t)$, $i=1,\dots,N$, $t=1,\dots,T$ is defined as 
\begin{align}
\label{eq:logreturns}
r_{i}{(t)} &= \log \left(\frac{c_i{(t)}}{c_i{(t-1)}}\right).
\end{align}
Our primary objects of study in this section are correlation matrices of log returns across rolling time windows of length $\tau$; here, we choose $\tau=120$ days. We standardise the log returns over such a window $[t-\tau+1,t]$ by defining ${R}_i(s) = [r_i(s) - \langle r_i \rangle] / \sigma(r_i)$ where $\langle . \rangle $ denotes the temporal average over the time window $[t-\tau+1,t]$ and $\sigma$ the associated standard deviation. The correlation matrix $\bm \Psi$ is then defined as follows: let $\bf{R}$ be the $N \times \tau$ matrix defined by ${R}_{is}={R}_i(s)$ with $i=1,\dots,N$ and $s=t-\tau+1,\dots,t$ and let
\begin{align}
\label{eq:corrmatrix}
\bm{\Psi}(t) = \frac{1}{\tau} {\bf{R}} {\bf{R}}^T.
\end{align}
Explicitly, individual entries describing the correlation behaviour between equities $i$ and $j$ are defined,
\begin{align}
\label{eq:rhodefn}
    \Psi_{ij}(t)=\frac{1}{\tau} \frac{\sum_{s=t-\tau+1}^t (r_i(s) - \langle r_i \rangle)(r_j(s) - \langle{r}_j \rangle)}{\left(\sum_{s=t-\tau+1}^t (r_i(s) - \langle r_i \rangle)^2 \right)^{1/2} \left( \sum_{s=t-\tau+1}^t (r_j(s) - \langle r_j \rangle)^2\right)^{1/2}},
\end{align}
for  $1\leq i, j\leq N$. We may analogously define the cross-correlation matrices for each individual sector by restricting $i$ and $j$ to be chosen from a set of indices corresponding to a particular sector. 

All entries $\Psi_{ij}$ lie in $[-1,1]$. $\bm\Psi$ is a symmetric positive semi-definite matrix with real and non-negative eigenvalues $\lambda_i(t)$, so we may order them as $\lambda_1 \geq \cdots \geq \lambda_N \geq 0$. As all diagonal entries of $\bm\Psi$ are equal to 1, the trace of $\bm\Psi$ is equal to $N$. Thus, we may normalise the eigenvalues by defining $\tilde{\lambda}_i = \frac{\lambda_i}{\sum^{N}_{j=1} \lambda_j}= \frac{\lambda_i}{N}$. 

Principal component analysis has been a corner stone in the analysis of dominant patterns in multivariate time series \cite{Jolliffe2011} and has been widely applied to financial data (see, for example, \cite{Fenn2011}). The eigenvectors ${\bf{v}}_i$ of the cross-correlation matrix $\bf{R}$, which we assume to be normalised throughout here,  capture directions of maximal variance of the data in a time period of length $\tau$, and the eigenvalues $\tilde{\lambda}_i$ capture how much of the observed variance of the data in that period can be described by the respective eigenvectors. In particular, $\tilde \lambda_i$ describes the proportion that the $i$th eigenvector ${\bf v}_i$ is able to reproduce the data. Hence if there are only a few eigenvalues of large magnitude, the data can be described by a linear combination of a few dominant eigenvectors. We are particularly interested in $\tilde{\lambda}_1(t)=\lambda_1(t)/N$ as a function of the rolling $\tau=120$-day window. Indeed, in the extreme case that $\tilde \lambda_1$ is close to $1$, the data can be described by the single mode ${\bf v}_1$, which we refer to as the {\em{market}}. Hence, if the temporal evolution of equities is dominated by a single mode, then all the variance in the data can be explained by ${\bf v}_1$, and there is no significant contribution of variance coming from other subspaces spanned by higher eigenvectors. In this sense we define $\tilde{\lambda}_1$ as a measure of the strength in collective correlations among a group of equities and as a proxy for a potential benefit of diversification.

\begin{figure*}
    \centering
    \includegraphics[width=\textwidth]{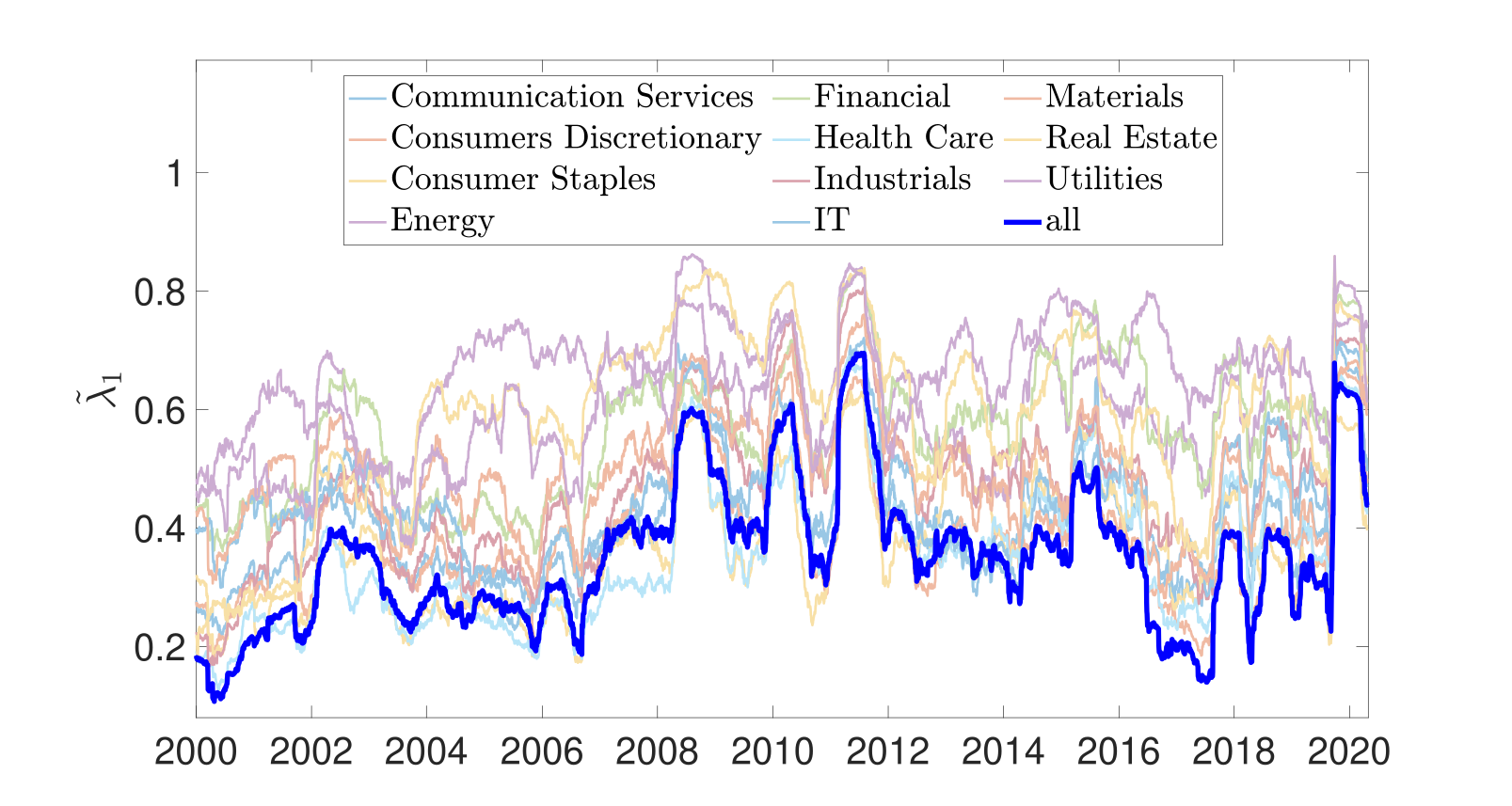}
    \caption{Normalised leading eigenvalue of the cross-correlation matrix as a function of time. Results are shown for the whole market consisting of all equities and for the 11 GICS sectors. One can see that collective correlations spike during market crises, such as the GFC and COVID-19. In addition, the collective correlations within individual sectors are consistently greater than that of the entire market. }
    \label{fig:Lambda1_sector}
\end{figure*}

Figure \ref{fig:Lambda1_sector} shows the evolution of the leading eigenvalue $\tilde \lambda_1(t)$ of the correlation matrix for all GICS sectors and the entire collection of equities, over the 20-year period we examine. There are several noteworthy findings. First, the leading eigenvalue attains large values during the two most prominent market crises, the global financial crisis (GFC) in 2008/2009  and the COVID-19 market crash in 2020. The GFC features three spikes in short succession commencing in late 2008 and the subsequent severe market responses in 2010 and 2011. By contrast, the COVID-19 market crash corresponds to one pronounced spike sustained for a period in early 2020. During bear markets and crises, the magnitude of the leading eigenvalue increases, often sharply, to large values - this heralds increased correlation between all underlying equities and less opportunity for successfully diversifying a portfolio returns stream.  Spikes of the leading eigenvalue can be explained by indiscriminate selling of risky assets (including equities) by both active and passive funds management businesses. Such spikes in the leading eigenvalue are associated with increased correlations among all underlying equities, and pronounced negative returns exhibited by equities with significant market beta. This suggests that during periods of large values of $\tilde \lambda_1$ diversification may not be beneficial as the overall systematic risk dominates over the unsystematic risk. During bull markets, for example during the extended period from 2016-2019, the normalised leading eigenvalue $\tilde \lambda_1$ can experience large fluctuations, however, the overall magnitudes are small, pointing to a lesser degree of correlations between equities. This lesser degree of correlations can be utilised by investors to diversify their portfolio.

Second, all sectors display broadly similar evolution over time regarding the peaks and troughs of their leading eigenvalue. Third, the degree of collectivity $\tilde \lambda_1$ is larger at all times when calculated from a cross-correlation matrix restricted to individual sectors than when calculated using all equities. This is consistent with our intuitive argument that diversification is more beneficial investing across sectors rather than within. Equities within a sector are more likely to be mutually correlated. An interesting observation is the absence of a spike in the leading eigenvalue around the time of the dot-com bubble in 2000/2001, in particular in the Information Technology sector. There are several possible explanations for this. First, most financial datasets spanning a significant period of time, such as ours, suffer from survivorship bias. Many technology-related companies went bankrupt during this period (including Pets.co, Webvan and 360Networks) and no longer exist within our dataset. Second, many companies that are generally thought of (and often classified) as Information Technology companies, may be classified in other GICS sector. One prominent example of this is Amazon's classification within the Consumer Discretionary sector. Factors such as these may have dampened the determination of equity collective behaviours (and the degree of severity) of the dot-com crisis, especially among the Information Technology sector.

Finally, we further investigate the extent of uniformity of the leading eigenvector ${\bf v}_1$ by introducing 
\begin{align}
\label{eq:normalisedip}
h(t) = \frac{|\langle {\bf v}_1,\bm{1}\rangle |}{ \|{\bf v}_1\| \| \bm{1} \|  },
\end{align}
where $\bm{1}=(1,1,\dots,1) \in \mathbb{R}^{\tilde N}$ and $\tilde N$ denotes the size of the underlying equities used to construct the cross-correlation matrix (\ref{eq:corrmatrix}). We remark that when the whole equity market is considered $ \| \bm{1} \| =N$ and if only a particular of the $M=11$ sectors is considered then $\| \bm{1} \|$ equals the size of that sector. Note that $h(t)\le 1$ with $h(t)=1$ for ${\bf v}_1={\bm 1}$. In this case, all equities carry the same amount of variance. This can be used to quantify the potential benefit of diversification: Increased values of $h(t)$ indicate increased interchangeability of equities and hence less opportunity for diversification or judicious selection of individual equities.   

\begin{figure*}
    \centering
    \includegraphics[width=\textwidth]{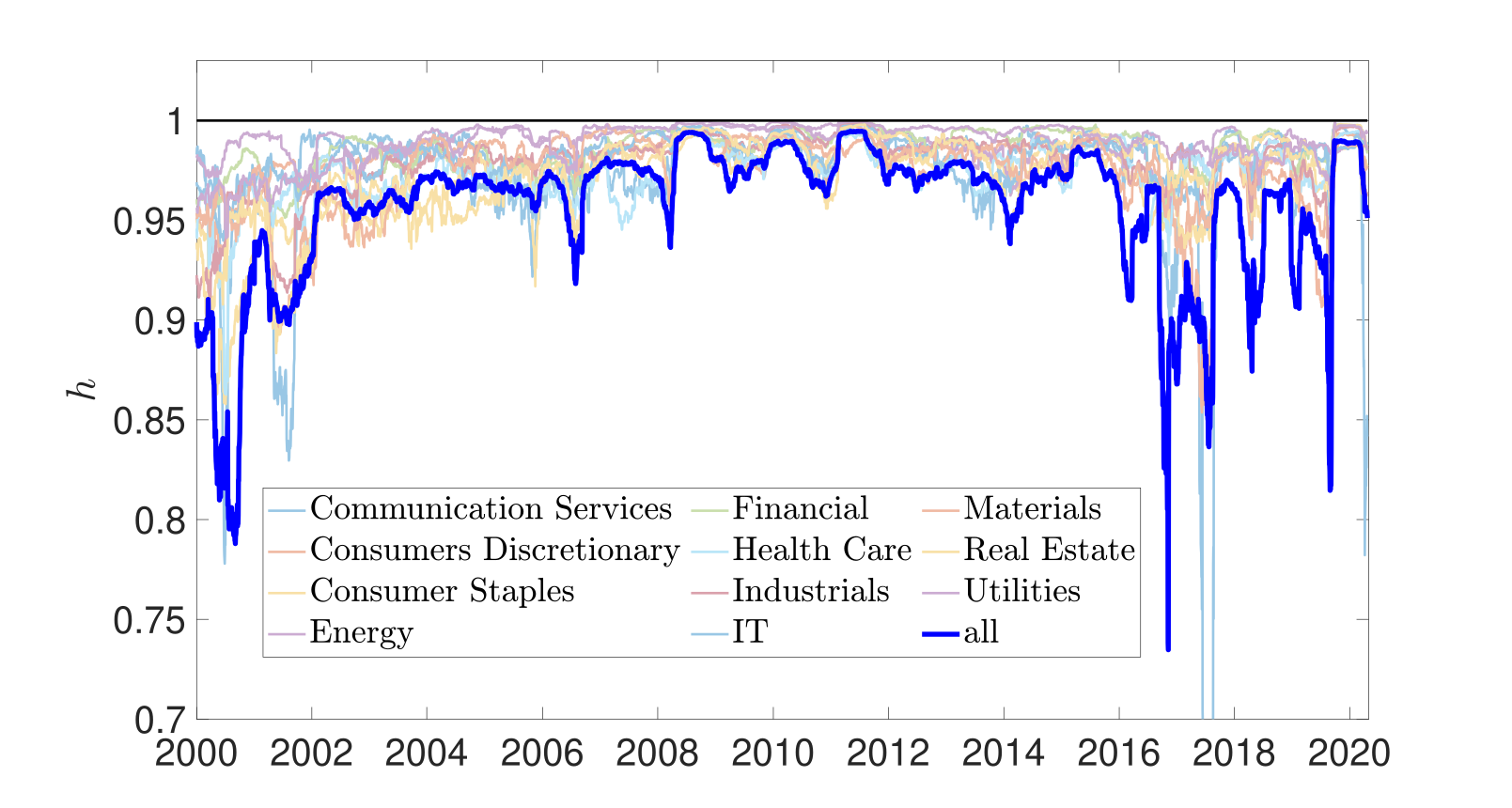}
    \caption{Uniformity of the leading eigenvector of the cross-correlation matrix as a function of time. Results are shown for the whole market consisting of all equities and for the 11 GICS sectors. Once again, the collective uniformity is greater for each individual sector than for the entire market, and uniformity is generally greater during market crises such as the GFC and COVID-19.}
    \label{fig:uniformity_sector}
\end{figure*}

In Figure \ref{fig:uniformity_sector}, we plot the uniformity measure $h(t)$ for each sector and for the entire market. The results are consistent with those shown in Figure \ref{fig:Lambda1_sector} for the degree of collectivity. As for the leading eigenvalue $\tilde {\lambda}_1$, the degree of uniformity $h(t)$ spikes during market crises (GFC and COVID-19), both for the individual sectors as well as for the entire market.  

We complement the spectral analysis of the cross-correlation matrix (\ref{eq:corrmatrix}) with a graph-theoretic view of the cross-correlation matrix over time. We view the correlation matrix as an adjacency matrix of a weighted graph to uncover the presence or absence of correlated sectors. 
Specifically, we consider a weighted graph with adjacency matrix $A$ with $A_{ij} = |C_{ij}|$. Unlike usual network-based community-finding algorithms, which are designed to identify communities purely from the structure of the adjacency matrix  \cite{Newman2018,vonLuxburg2007,Fenn2011,Fortunato2010}, we assume here that the given sectors predetermine the communities \textit{a priori}. The graph-theoretic diagnostics are then used to quantify the strength of the partition of the graph into those fixed sectors.

We study in particular the (rolling) modularity associated with the partition of the graph defined by the sectors. Modularity measures the difference between the observed number of (weighted) edges within a sector and the expected number of edges if they were randomly assigned \cite{Newman2018}. Treating each individual asset as a vertex, its degree is defined as $k_i=\sum_{j=1}^N A_{ij}$, while the total number of edges (counted by weight) of the graph is $e=\frac12 \sum_{i=1}^N k_i$. Denoting by ${\mathcal{S}}_m$ the set of equities which make up the $m$th sector the modularity is defined as 
\begin{align}
Q= \frac{1}{2e} \sum_{m=1}^M\sum_{i,j\in {\mathcal{S}}_m} \left( A_{ij}-\frac{k_ik_j}{2e}\right).
\label{e.Q}
\end{align}
As elsewhere in this section, we compute and study $Q$ on a rolling $\tau=120$-day basis.

\begin{figure*}
    \centering
    \includegraphics[width=\textwidth]{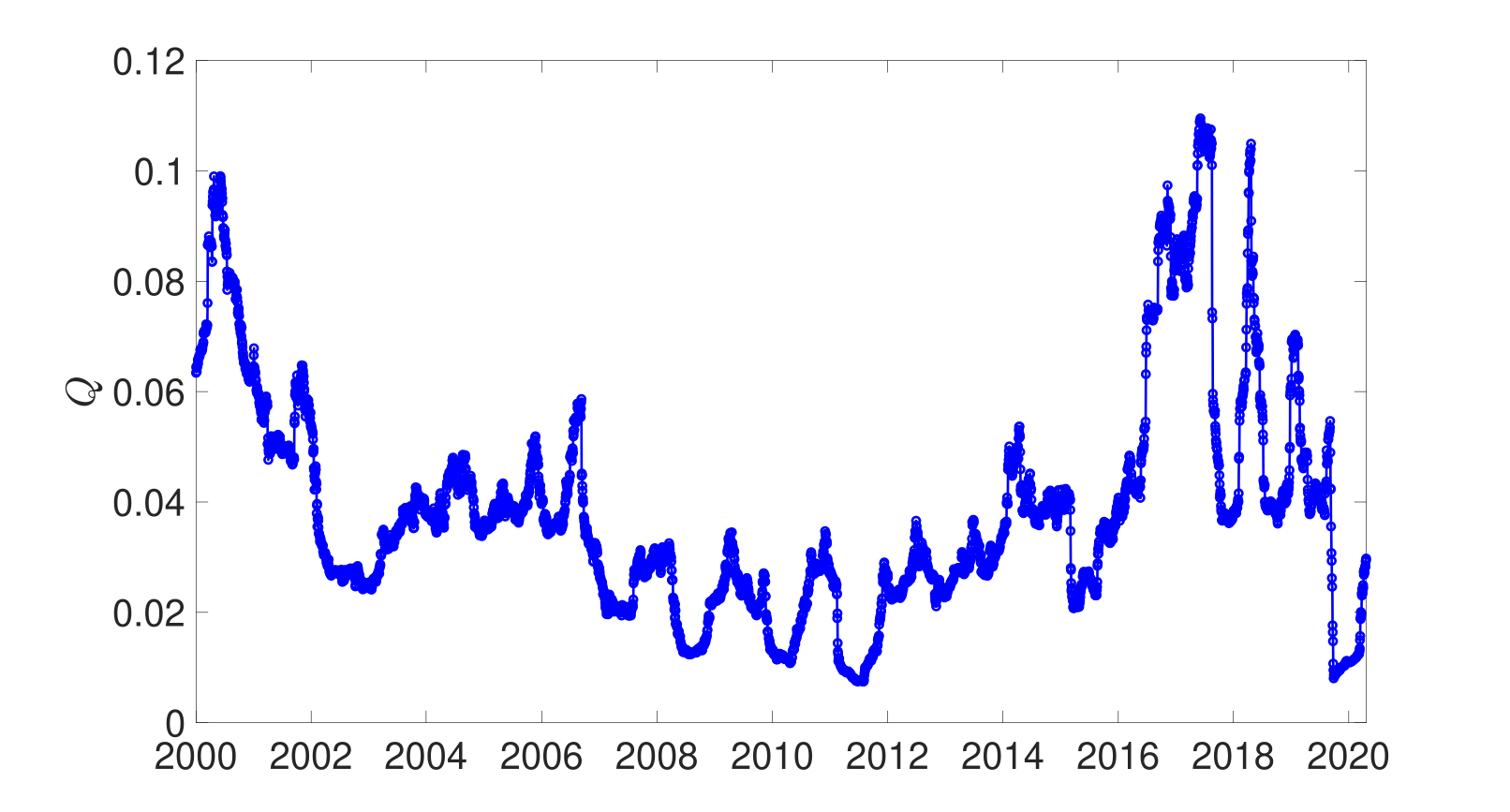}
    \caption{Modularity $Q$ for the partition consisting of the 11 the GICS sectors (online blue). Financial crises are identified as times with small modularity, indicating that sectors cease to constitute independent sets of equities that are more correlated among themselves than with other sectors. }
    \label{fig:modularity}
\end{figure*}

In Figure \ref{fig:modularity}, we show the evolution in modularity $Q$ for the partition defined by the GICS sectors. Consistent with the degree of collectivity $\tilde \lambda_1$ and the uniformity measure $h(t)$, modularity clearly identifies financial crises as events with small modularity $Q$, indicating that in times of financial crises sectors cease to constitute independent sets of equities which are more correlated with each other than with equities from other sectors. In fact, there are only four events in time where the level of modularity drops below 0.018 - the three troughs corresponding to the GFC between 2008 and 2012, and the COVID-19 market crash in 2020. In contrast, modularity is highest during the mid-2000s and during the equity bull market of 2016-2019. As for the normalised leading eigenvalue $\tilde\lambda_1$ (cf. Fig.~\ref{fig:Lambda1_sector}), the modularity experiences large fluctuations albeit with values significantly larger than those experienced during financial crises. To test if sectors constitute reasonable communities in the sense that there are more (weighted) edges linking equities within a sector than what one expects from a random allocation of edges, we calculated the average modularity over 500 random allocations of equities to 11 random groupings of the same size as the original sectors.  The averaged modularity of this ensemble of randomised sectors is of the order of 5 times smaller than the modularity of the actual market with its sectors (not shown). Interestingly though the temporal evolution experiences the same troughs and spikes as the modularity $Q$ shown in Figure~\ref{fig:modularity}.

We have shown in this Section that the normalised leading eigenvalue of the cross-correlation matrix $\tilde \lambda_1(t)$, the equity uniformity measure $h(t)$ and the modularity $Q(t)$ have very similar signatures. All these diagnostics can be used to identify collective behaviour of the market and hence to assess the potential benefit of diversification. In the following Section we will use the normalised leading eigenvalue as our proxy for a diversification benefit.

We remark on the choice of the time window length $\tau=120$ used to construct the cross-correlation matrix (\ref{eq:corrmatrix}). The choice of this parameter is a delicate balance between excessive and insufficient smoothing. If the value of $\tau$ is chosen too large, the level of smoothing will be excessive, and we may be unable to identify abrupt changes in the correlation structure. A prime example of this is the COVID-19 market crash, which was extremely severe, albeit quite brief. Alternatively, if $\tau$ is chosen too small, we may erroneously interpret short-term transient noisy events as meaningful changes of the correlation structure. The size of the smoothing window varies significantly in the literature, ranging from windows of 3 months to 2 years of trading data \cite{JamescryptoEq,Fenn2011}, depending on the time-scales of interest. Here we are interested in both abrupt market changes developing over a few months as well as longer-term structural shifts in correlation patterns, motivating the compromising value of $\tau=120 $ days, i.e. 6 months. 

We use the whole data set comprised of several periods of bear and bull markets and did not stratify the data according to different market dynamics. First, this allows us to investigate the long-term implications of equity portfolio diversification strategies, which consists of bull and bear market periods. Second, given that market dynamics varies significantly over time (and no two market crises are ever the same), the optimal portfolio structure of a previous period of economic crisis or stability, may not be ideal in a similarly-themed future period. Finally, it may be difficult for retail investors to anticipate changes in equity market dynamics, and restructure their portfolios based on expected equity market performance. Accordingly, we study optimal diversification strategies for equity market investors who may lack the resources, information or interest to constantly re-balance their equity sector exposure based on market sentiment. 

\section{Portfolio sampling}
\label{sec:portfoliosampling}

We now perform an extensive sampling procedure to explore how diversification benefits depend on the number of equities held in the portfolio and on the number of sectors from which to choose those equities. To quantify the potential diversification benefit we choose here, motivated by the results obtained in Section~\ref{sec:marketstructure}, the degree of collective behaviour of the market $\tilde \lambda_1(t)$. We study the diversification benefits of portfolios that consist of $mn$ equities such that $n$ equities are drawn from $m$ separate sectors. Both the individual equities and the sectors are drawn randomly and independently with uniform probability. We draw $D=500$ portfolios for each combination $(m,n)$.

To quantify the potential diversification benefit for a portfolio consisting of $mn$ equities, we determine the $mn\times mn$ correlation matrix $\bm \Psi$ for each draw and calculate the associated normalised eigenvalues $\tilde{\lambda}_{m,n}(t)$. We again use a rolling time window of length $\tau=120$ days when determining the cross-correlation matrix. For each combination $(m,n)$ of number of sectors $m$ and number of equities per sector $n$ we record the 5th percentile, 50th percentile (median) and the 95th percentile of the $D$ values of $\tilde{\lambda}_{m,n}(t)$. These are denoted by $\tilde{\lambda}_{m,n}^{0.05}(t), \tilde{\lambda}_{m,n}^{0.50}(t)$ and  $\tilde{\lambda}_{m,n}^{0.95}(t)$, respectively.

\begin{figure*}
\centering
\includegraphics[width=0.8\textwidth]{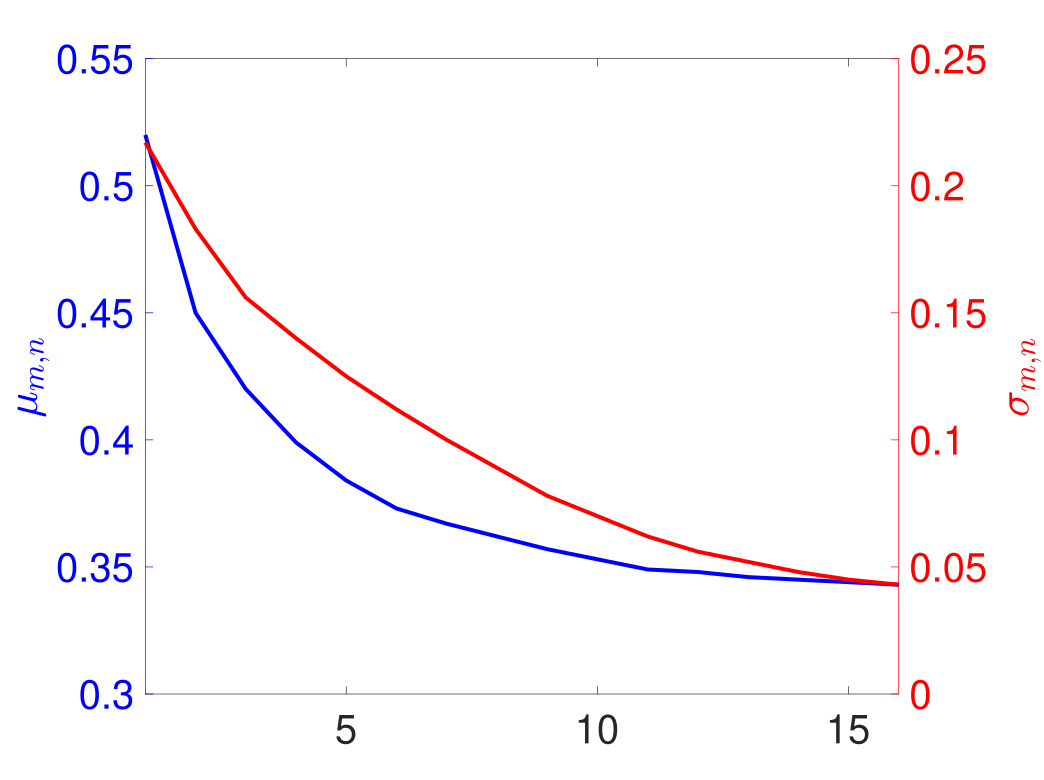}
\caption{Median $\mu_{m,n}$ and spread $\sigma_{m,n}$ of the normalised eigenvalue for the greedy path depicted in Tables~\ref{tab:samplemeans} and \ref{tab:samplevariances}, starting from the smallest $(2,2)$ portfolio and ending up at the largest $(10,9)$ portfolio. We see that the diversification benefit of larger portfolios (measured by the median of the first eigenvalue $\mu_{m,n}$) exhibits sharply diminishing returns after about 10 steps of increasing $m$ and $n$.}
\label{fig:greedypath}
\end{figure*}

We introduce the temporal mean of the median of the normalised eigenvalues 
\begin{align}
  \mu_{m,n}= \frac{1}{T-\tau+1} \sum_{t=\tau}^T \tilde{\lambda}_{m,n}^{0.50}(t)
\end{align}
as a measure of the diversification benefit of a portfolio with $n$ stocks in each of $m$ sectors. Table~\ref{tab:samplemeans} records $\mu_{m,n}$ for portfolios with $2\le m \le  10$ and $2\le n\le 9$. In the following we denote by $(m,n)$ a portfolio with $n$ equities chosen from $m$ separate sectors. As expected, the diversification benefit is seen to be smallest for the smallest portfolio $(2,2)$ consisting of 4 equities and is largest for the largest portfolio $(10,9)$ consisting of 90 equities. Table~\ref{tab:samplemeans} reveals that if we want to keep the total number $mn$ of equities contained in a portfolio constant, we have $\mu_{m,n} < \mu_{n,m}$ showing that it is more beneficial to diversify across sectors than within sectors.  We can fix the number of sectors $m$ and increase $n$ to see a reduction in the magnitude of $\mu_{m,n}$ implying as expected a larger diversification benefit. Similarly, we can fix the number of equities form each sector and increase the number of sectors $m$. The decrease of $\mu_{m,n}$ is stronger here for the increase of the number of sectors when compared to the previous scenario where the number of equities is varied, again pointing to the fact that diversifying across sectors is more beneficial than within sectors. We show in Table~\ref{tab:samplemeans} a greedy strategy (online red) where, starting at the smallest portfolio $(2,2)$ and ending at the largest portfolio $(10,9)$ we aim to decrease the value of $\mu_{m,n}$ by either increasing the number of sectors $m$ to choose from or the number of equities $n$ to be chosen from each of those sectors. The greedy path is shown in Fig.~\ref{fig:greedypath}. It is seen that the median $\mu_{m,n}$ saturates and that not much is gained by increasing the number of sectors from 9 to 10. The question we ask in this Section is whether we can find a portfolio which results in a comparable diversification benefit to the largest $(10,9)$ portfolio but which contains significantly less number of equities? Since there is no significant difference in our measure for the diversification benefit $\mu_{m,n}$ for the $(9,9)$ and the $(10,9)$ portfolios we restrict our analysis from now on to portfolios with a maximum of 9 sectors, with the $(9,9)$ portfolio being the most diversified portfolio. The smallest portfolio which has a value of $\mu_{m,n}$ comparable to the minimal value of the $(10,9)$ and $(9,9)$ portfolios is identified to be a $(9,4)$ portfolio. Using hierarchical clustering we will show below that indeed the $(9,4)$ portfolio with a total of 36 equities behaves close to the most diversified $(9,9)$ portfolio with a total of 81 equities.

\begin{figure*}
\centering
\begin{subfigure}[b]{0.99\textwidth}
\includegraphics[width=\textwidth]{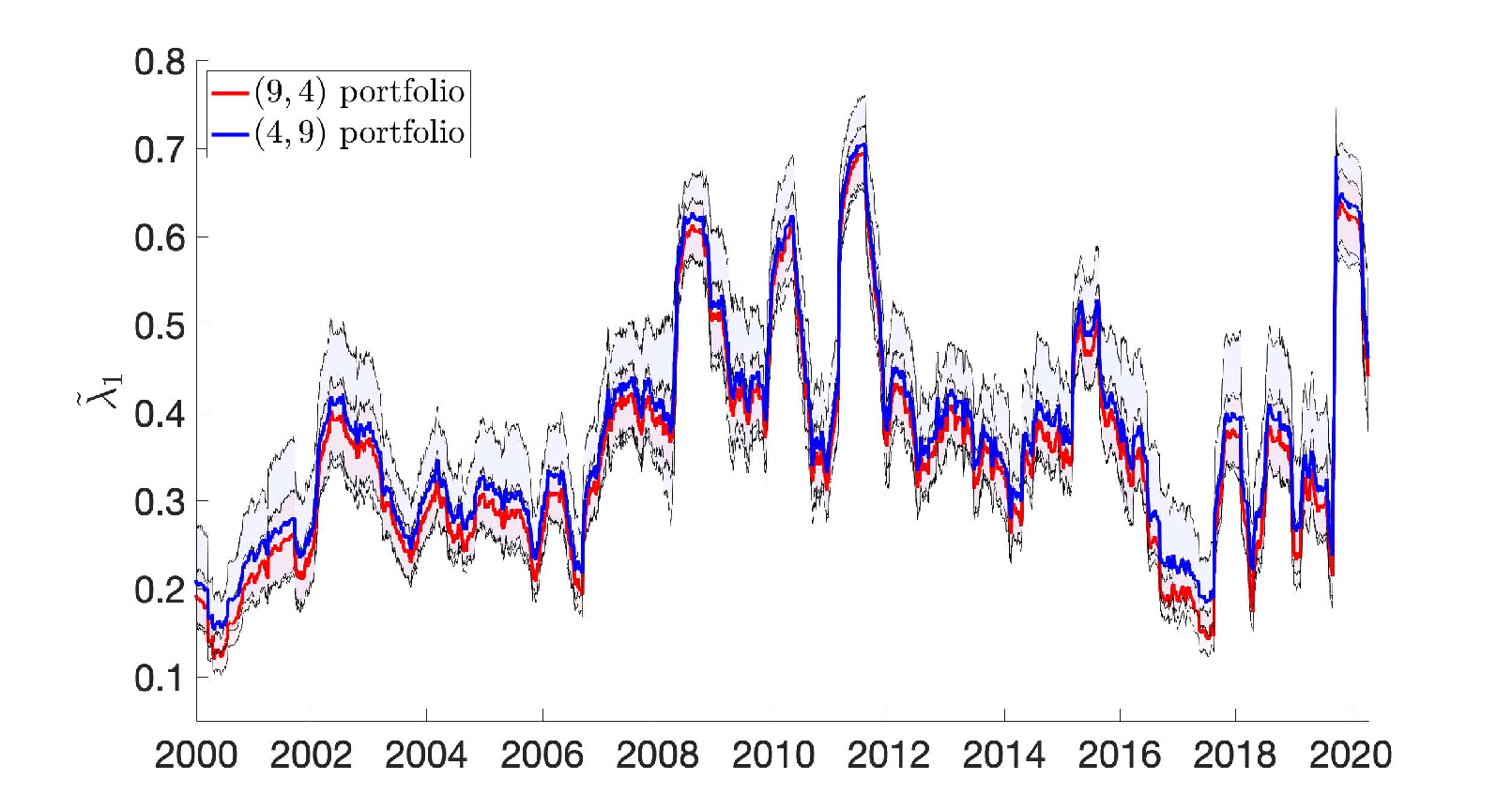}
\caption{}
\label{fig:94_49}
\end{subfigure}
\begin{subfigure}[b]{0.99\textwidth}
\includegraphics[width=\textwidth]{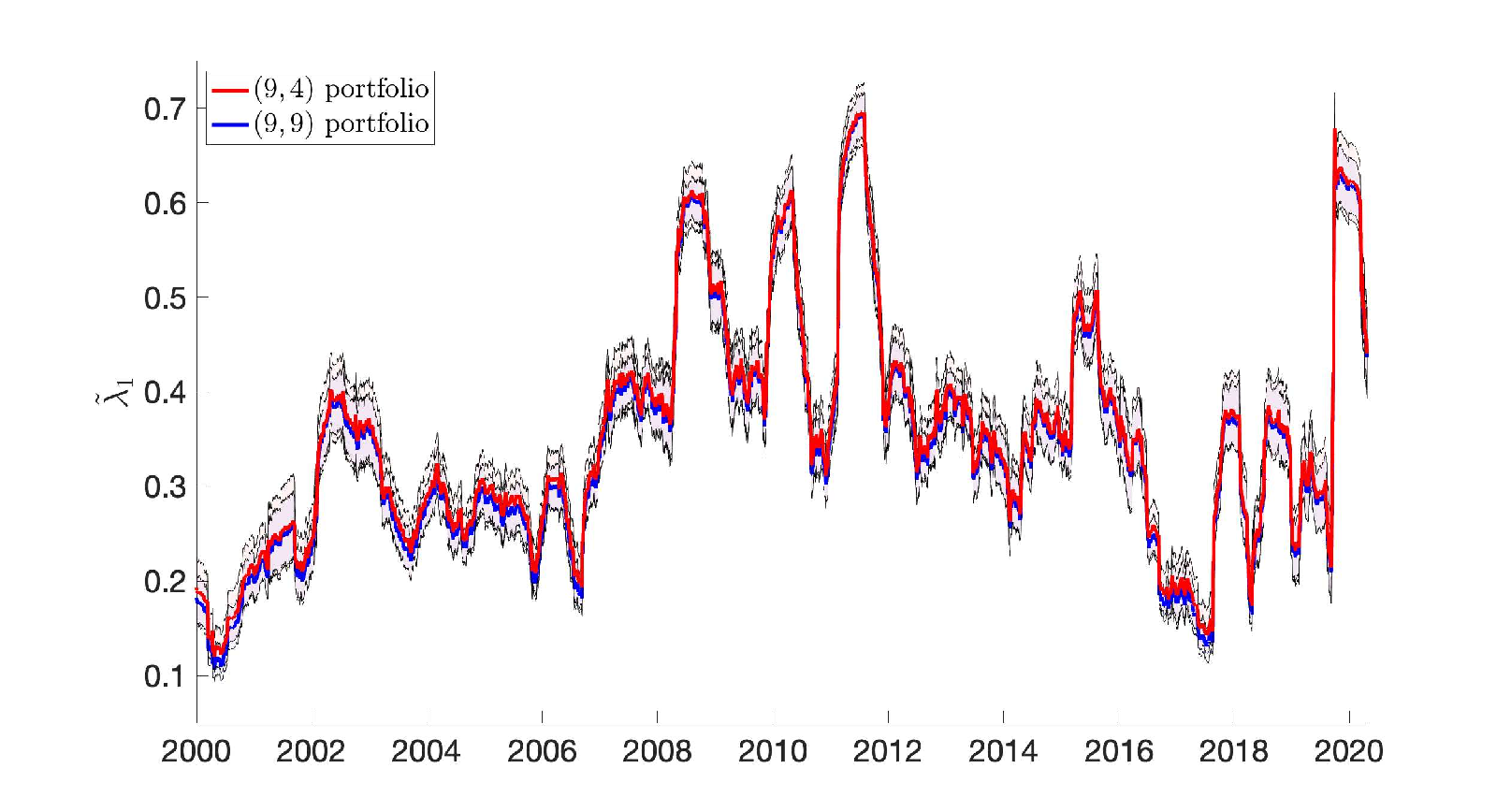}
\caption{}
\label{fig:94_99}
\end{subfigure}
\caption{Median of the normalised eigenvalues of the cross-correlation function $\tilde{\lambda}_{m,n}^{0.50}(t)$, together with the 5th and 95th percentiles $\tilde{\lambda}_{m,n}^{0.05}(t)$ and $\tilde{\lambda}_{m,n}^{0.95}(t)$ for several portfolios. (a): $(9,4)$ portfolio (online blue) and $(4,9)$ portfolio, both containing $36$. (b): $(9,4)$ portfolio and largest $(9,9)$ portfolio. We see that the figures for the (9,4) and (9,9) portfolio are much more similar than that of the (9,4) and (4,9) portfolio, indicating the considerable diversification benefit of the (9,4) portfolio and the advantage of holding a greater number of sectors than stocks per sectors.}
\label{fig:diffsamples}
\end{figure*}

To explore the $(9,4)$ portfolio in more detail and how it compares to portfolios of the same size such as a $(4,9)$ portfolio as well as to the most diversified $(9,9)$ portfolio we show in Figure~\ref{fig:diffsamples} the temporal evolution of $\tilde{\lambda}_{m,n}^{0.50}$. It is clearly seen in Figure~\ref{fig:diffsamples}a that the $(9,4)$ portfolio exhibits smaller values of $\tilde{\lambda}_{m,n}^{0.50}$ compared to the $(4,9)$ portfolio with the same number of total equities held at all times, independent of whether the market experiences a financial crisis or a bull market. Moreover, it is seen that the spread of the $(9,4)$ portfolio, as measured by the distance between the 5th and the 95th percentile curves is smaller for the $(9,4)$ portfolio. Remarkably, as seen in Figure~\ref{fig:diffsamples}b, the curves of $\tilde{\lambda}_{m,n}^{0.50}(t)$ of the $(9,4)$ portfolio closely resembles that of the largest $(9,9)$ portfolio, with comparable spread. This shows that the diversification benefit of the smaller $(9,4)$ portfolio is very similar to the much larger $(9,9)$ portfolio. 

The previous discussion was centred around the average behaviour of a portfolio with a specified number of sectors and equities per sectors. For investors it is of paramount importance to know if the average behaviour is typical. If this is not the case then the diversification benefit will strongly depend on the particular choice of the equities picked form each sector. We expect that the variance will be larger in smaller portfolios, with a maximum at the $(2,2)$ portfolio, and will decrease with increasing number of equities held, with a minimum variance for the largest $(9,9)$ portfolio. To quantify this we look at the average spread defined by
\begin{align}
    \sigma_{m,n}= \frac{1}{T-\tau+1} \sum_{t=\tau}^T \tilde{\lambda}_{m,n}^{0.95}(t) - \tilde{\lambda}_{m,n}^{0.05}(t).
\end{align}
The difference between the 5th and 95th percentile of a distribution corresponds, under the assumption of Gaussianity, to approximately 1.96 times the underlying standard deviation. We record $\sigma_{m,n}$ in Table \ref{tab:samplevariances}. As for the average $\mu_{m,n}$ we observe that $\sigma_{m,n} < \sigma_{n,m}$, implying that to construct a portfolio of $mn$ equities it is more beneficial to increase the number of sectors than the number of equities held per sector. Similarly, the decrease in the spread is more pronounced when for a fixed number of equities we increase the number of sectors to choose from than for the case when for a fixed number of sectors the number of equities per sector is increased. 

\begin{table*}
\centering
\begin{tabular}{c|rrrrrrrr}
  \hline
 & \multicolumn{8}{c}{Number of equities per sector} \\
  \hline
Number of sectors & 2 & 3 & 4 & 5 & 6 & 7 & 8 & 9 \\ 
  \hline
  2 &\textcolor{red}{0.520} & 0.480 & 0.460 & 0.450 & 0.440 & 0.430 & 0.420 & 0.440 \\ 
  3 & \textcolor{red}{0.450} & \textcolor{red}{0.420} & 0.410 & 0.406 & 0.397 & 0.396 & 0.388 & 0.390 \\ 
  4 & \textcolor{red}{0.420} & \textcolor{red}{0.399} & 0.393 & 0.386 & 0.378 & 0.375 & 0.373 & 0.373 \\ 
  5 & 0.400 & \textcolor{red}{0.384} & 0.376 & 0.369 & 0.368 & 0.365 & 0.363 & 0.362 \\ 
  6 & 0.389 & \textcolor{red}{0.373} & 0.368 & 0.363 & 0.360 & 0.359 & 0.356 & 0.354 \\ 
  7 & 0.379 & \textcolor{red}{0.367} & \textcolor{red}{0.362} & 0.358 & 0.355 & 0.352 & 0.351 & 0.351 \\ 
  8 & 0.373 & \textcolor{red}{0.362} & \textcolor{red}{0.357} & 0.354 & 0.351 & 0.350 & 0.348 & 0.348 \\ 
  9 & 0.368 & 0.358 & \textcolor{red}{0.353} & \textcolor{red}{0.349} & \textcolor{red}{0.348} & 0.347 & 0.345 & 0.345 \\ 
    10 & 0.364 & 0.355 & 0.35 & 0.348 & \textcolor{red}{0.346} & \textcolor{red}{0.345} & \textcolor{red}{0.344} & \textcolor{red}{0.343} \\ 
   \hline
\end{tabular}
\caption{Average $\mu_{m,n}$ of the median normalised eigenvalue $\tilde{\lambda}_{m,n}^{0.50}(t)$ for different pairs of $m$ sectors and $n$ equities per sectors. In red, we display a greedy strategy how to reduce the value of $\mu_{m,n}$ (implying an increase the overall diversification benefit) by gradually increasing the portfolio size, starting from the smallest portfolio $(2,2)$.}
\label{tab:samplemeans}
\end{table*}

\begin{table*}[ht]
\centering
\begin{tabular}{c|rrrrrrrr}
  \hline
 & \multicolumn{8}{c}{Number of equities per sector} \\
  \hline
Number of sectors & 2 & 3 & 4 & 5 & 6 & 7 & 8 & 9 \\ 
  \hline
2 & \textcolor{red}{0.217} & 0.210 & 0.203 & 0.202 & 0.199 & 0.195 & 0.154 & 0.155 \\ 
  3 & \textcolor{red}{0.183} & 0.169 & 0.159 & 0.157 & 0.151 & 0.150 & 0.147 & 0.148 \\ 
  4 & \textcolor{red}{0.156} & 0.144 & 0.138 & 0.132 & 0.125 & 0.127 & 0.121 & 0.124 \\ 
  5 & \textcolor{red}{0.140} & 0.127 & 0.118 & 0.114 & 0.109 & 0.106 & 0.106 & 0.100 \\ 
  6 & \textcolor{red}{0.125} & \textcolor{red}{0.112} & 0.104 & 0.101 & 0.097 & 0.093 & 0.090 & 0.087 \\ 
  7 & 0.116 & \textcolor{red}{0.100} & 0.095 & 0.087 & 0.083 & 0.079 & 0.078 & 0.076 \\ 
  8 & 0.102 & \textcolor{red}{0.089} & 0.081 & 0.076 & 0.071 & 0.069 & 0.068 & 0.065 \\ 
  9 & 0.094 & \textcolor{red}{0.078} & \textcolor{red}{0.070} & 0.066 & 0.062 & 0.059 & 0.057 & 0.054 \\ 
  10 & 0.085 & 0.070 & \textcolor{red}{0.062} & \textcolor{red}{0.056} & \textcolor{red}{0.052} & \textcolor{red}{0.048} & \textcolor{red}{0.045} & \textcolor{red}{0.043} \\ 
   \hline
\end{tabular}
\caption{Average spread $\sigma_{m,n}$ of the median normalised eigenvalues for different pairs of $m$ sectors and $n$ equities per sectors. In red, we display a greedy strategy how to reduce the value of $\sigma_{m,n}$ (implying a smaller dependency on the portfolio selection) by gradually increasing the portfolio size, starting from the smallest portfolio $(2,2)$.}
\label{tab:samplevariances}
\end{table*}

We now address the question which portfolio combinations $(m,n)$ share the most similar evolution in their collective dynamics? This allows us to determine the smallest portfolio which has a comparable diversification benefit to the most diversified $(9,9)$ portfolio. To tackle this question, we perform hierarchical clustering on the distance metric 
\begin{align}
\label{eq:distanced}
    d((m,n),(m',n'))= \frac{1}{T-\tau+1} \sum_{t=\tau}^T| \tilde{\lambda}_{m,n}^{0.50}(t) - \tilde{\lambda}_{m',n'}^{0.50}(t)|
\end{align}

which quantifies the average absolute difference between the median eigenvalues of two portfolios $(m,n)$ and $(m',n')$. This results in a $64 \times 64$ distance matrix for $2\le m,n\le 9$. Given the relatively high dimensionality of the data, we choose the Manhattan distance over other alternatives such as the Euclidean distance. However, we checked that the key findings remain unchanged when using the Euclidean distance instead. Once the distance matrix has been formed, we apply hierarchical clustering to determine which portfolio combinations share the most similarity in their collective behaviour evolution. Hierarchical clustering is a convenient tool to reveal proximity between different elements of a collection. Here, we perform agglomerative hierarchical clustering based on the average-linkage criterion \cite{Mllner2013}. The algorithm works in a bottom-up manner, where each portfolio combination starts in its own cluster, and pairs of clusters are merged as one traverses up the hierarchy. Given the high transaction cost investors may face when holding larger portfolios, we wish to identify the smallest (or best value) portfolios which provide the greatest risk reduction relative to the number of equities held. As in Section~\ref{sec:marketstructure} we compute distances $d((m,n),(m',n'))$ over the entire period, rather than stratifying according to different macroeconomic characteristics to addresses `all weather' risk mitigation of risk reduction across a range of market scenarios. 

\begin{figure*}
    \centering
    \includegraphics[width=\textwidth]{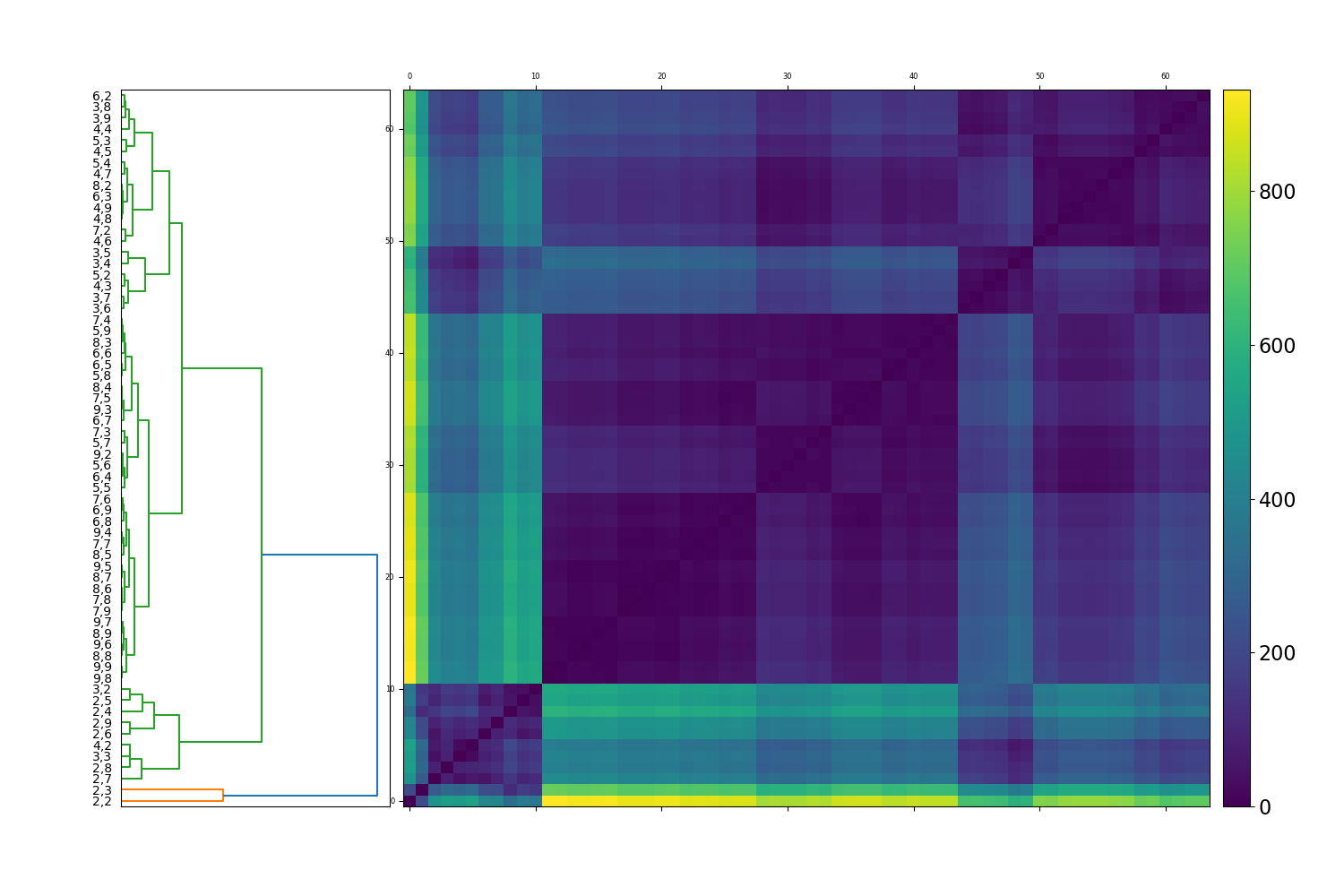}
    \caption{Hierarchical clustering between pairs $(m,n)$ of different portfolio structures, using the $L^1$ metric between median functions (\ref{eq:distanced}). Two outliers portfolios (the smallest size) are revealed, and then three subclusters of the majority collection. Of greatest interest is the dense dark partition of high similarity ranging from (7,6) to (9,8). This contains both the (9,4) and (9,9) portfolios, revealing that a 36-equity portfolio of 9 sectors and 4 equities per sector attains near-identical diversification benefit as the largest possible portfolio. This may help confirm the financial adage that 30-40 stocks may provide sufficient diversification, and strongly suggest the benefit of holding this structure of stock portfolio for a capacity-limited investor.}
    \label{fig:lambda1_dend}
\end{figure*}

The resulting dendrogram from this analysis is shown in Figure \ref{fig:lambda1_dend}. Clusters of similar evolution in their collective behaviour are identified as blue blocks along the anti-diagonal. The corresponding portfolio combinations are shown on the far left of Figure ~\ref{fig:lambda1_dend}. The darker blue colouring for any respective square block corresponds to less distance between evolutionary paths, and a higher degree of affinity.

The dendrogram exhibits 3 primary subclusters and a small outlier cluster. The outlier cluster (orange leaves) consists only of portfolios $(2,2)$ and $(2,3)$ which contain just 4 and 6 equities respectively and which provide the least diversification benefit (cf. Table~\ref{tab:samplemeans}). Directly above the outlier cluster is a subcluster of 9 relatively small portfolios ranging from $(3,2)$ to $(2,7)$ on the left-side panel. Excluding the outliers, this subcluster provides the least diversification benefit to an investor. The largest portfolio in this cluster is portfolio $(2,9)$ with 18 equities and the smallest portfolio is portfolio $(3,2)$ with only 6 equities. Both portfolios exhibit similar temporal evolution in terms of the median eigenvalues and hence similar levels of risk reduction. This confirms again that it is more advantageous to increase the number of sectors to construct a portfolio than the actual number of equities held.

The predominant subcluster in Figure \ref{fig:lambda1_dend} spans (according to the labels on the left-hand side) portfolios $(7,4)$ to $(9,8)$. It can be further subdivided into two subclusters. The first subcluster consists of portfolios ranging from $(7,4)$ to $(5,5)$. The largest portfolio in this cluster is portfolio $(5,9)$ with 45 equities and the smallest portfolio is portfolio $(7,3)$ with 21 equities, providing the same level of risk reduction. Again, it is seen that diversifying across sectors is much more effective in terms of diversification benefits than simply increasing the number of equities. The other of the two subclusters contains cluster (according to the left-hand side panel) $(7,6)$ to $(9,8)$. This subcluster contains portfolios which behave the most similar in terms of their median eigenvalues as seen by the dark blue colour. This cluster contains the largest portfolio $(9,9)$ and the smallest portfolio is our designated portfolio $(9,4)$. This has several important implications for equity-based portfolio management. First, there is an old adage in financial markets that 30-40 equities are sufficient for diversification and elimination of unsystematic portfolio risk. The $(9,4)$ portfolio, composed of 36 equities nicely fits into this range. Second, it is of great relevance to retail and cost-conscious investors, that a $(9,4)$ portfolio provides a nearly identical diversification benefit to a $(9,9)$ portfolio, more than twice its total size.

\section{Conclusion}
\label{sec:conclusion}

We have used spectral and graph-theoretical characteristics of the cross-correlation matrix of the log returns of equities in the US market from 2000 until 2020 to quantify the collective behaviour of equities over time as a diagnostics for potential diversification benefits in terms of identifying the dominance of systematic risk over unsystematic risk. We found that the leading eigenvalue, a uniformity measure and modularity can all be used to detect dominant collective behaviour in the market such as the GFC and the COVID-19 crisis as well as identify bear markets as encountered during the period from 2016-2019. We then studied the properties of random portfolios of a specific size. A major takeaway from our portfolio sampling and hierarchical clustering analysis is the identification of a best value `all weather' portfolio consisting of choosing 4 equities from each of 9 sectors, totalling 36 equities. The sampling procedure and respective dendrogram highlight that this portfolio provides comparable reduction in unsystematic risk to the largest and most diversified portfolio consisting of choosing 9 equities from each of 9 sectors, totalling 81 equities. The findings in this paper highlight optimal equity sector diversification strategies during a 20 year period which includes multiple periods of economic crisis, as well as periods of stability, and hence provide guidance for portfolio constructions in an `all weather' environment which is agnostic to the current macroeconomic environment. We verified that the actual choice of which sectors and which equities to choose from is not important in terms of risk reduction and the optimal $(9,4)$ portfolios exhibit very little spread, again comparable to the spread incurred by the largest $(9,9)$ portfolio. This supports the widely known rule of thumb that a portfolio consisting of 30-40 equities is sufficient in reducing unsystematic risk. Our results demonstrate that there is significantly greater benefit in diversifying equity portfolios across sectors than within sectors and a $(9,4)$ portfolio provides significantly larger risk reduction than, for example, a $(4,9)$ portfolio of equal total size. Reassuringly, for the optimal $(9,4)$ portfolio we found that the risk reduction does not depend strongly on the actual choice of sectors and equities in a long 20 year investment period.

There are several avenues of potential future research. First, it would be interesting to consider a market consisting of more than a single asset class and to include asset classes such as fixed income, currencies, commodities, cryptocurrencies and other alternative asset classes. In particular, it would be interesting to see if the graph-theoretic approach is able to identify separate community behaviours based on different asset classes, and if these could be further broken down into underlying constituent groupings (such as equity sectors). Similarly, it would be interesting to extend the portfolio sampling to include other asset classes. It is possible (and quite likely) that including more asset classes is conducive in the diversification of portfolios and reduces the tendency for correlated collective portfolio behaviour. Second, one could study similar phenomena to that explored in this paper in different geographies. It is possible that in some countries, market dynamics may be more or less correlated than that of the US equity market. Third, one could employ different association measures under Pearson correlation, including parametric approaches that explicitly take into account the heavy tails of financial returns. Finally, one could extend the portfolio sampling procedure to consider portfolio returns, in addition to risk. This paper specifically deals with the concept of portfolio diversification from the standpoint of reducing collective behaviours. If one were to consider the returns (in addition to the risk) in various portfolio settings, this could potentially be of great interest to the community of financial market researchers.


\appendix
\section{Equity securities}
\label{sec:equity_securities}

\begin{tabular}{lll}
\toprule
Sector &   Ticker & Name\\
\midrule
Communication Services &  ATVI  & Activision Blizzard \\
Communication Services &  T  & AT\&T  \\
Communication Services &  CMCSA  & Comcast Corp. \\
Communication Services &  DISH  & Dish Network Corp. \\
Communication Services &  EA   & Electronic Arts \\
Communication Services &  IPG  & IPG Photonics \\
Communication Services &  OMC  & Omnicom Group \\
Communication Services &  TTWO  & Take-two Interactive Software \\
Communication Services &  VZ  & Verizon Communications \\
Communication Services &  DIS   & Walt Disney Company \\
Consumer Discretionary &  AMZN  & Amazon.com \\
Consumer Discretionary &  AZO  & AutoZone Inc. \\
Consumer Discretionary &  F  & Ford Motor Co. \\
Consumer Discretionary &  GPS & Gap Inc. \\
Consumer Discretionary &  GPC  & Genuine Parts Company \\
Consumer Discretionary &  HRB  & H\&R Block \\
Consumer Discretionary &  HOG & Harley-Davidson \\
Consumer Discretionary &  HD  & Home Depot Inc. \\
Consumer Discretionary &  KSS & Kohl's Corp. \\
Consumer Discretionary &  LB   & Bath \& Body Works \\
Consumer Discretionary &  LEG  & Leggett \& Platt \\
Consumer Discretionary &  LEN  & Lennar Corp. \\
Consumer Discretionary &  BBY  & Best Buy Co. \\
Consumer Discretionary &  LOW  & Lowe's Cos Inc. \\
Consumer Discretionary &  MCD   & McDonald's Corp. \\
Consumer Discretionary &  MGM  & MGM Resorts International \\
Consumer Discretionary &  MHK  & Mohawk Industries \\
Consumer Discretionary &  NKE & Nike Inc. \\
Consumer Discretionary &  JWN  & Nordstrom Inc. \\
Consumer Discretionary &  ORLY  & O'Reilly Automotive Inc. \\
Consumer Discretionary &  PHM  & PulteGroup Inc. \\
Consumer Discretionary &  PVH  & PVH Corp. \\
Consumer Discretionary &  RL  & Ralph Lauren Corp. \\
Consumer Discretionary &  BKNG  & Booking Holdings Inc. \\
Consumer Discretionary &  ROST  & Ross Stores Inc. \\
Consumer Discretionary &  RCL  & Royal Caribbean Cruises Ltd. \\
Consumer Discretionary &  SBUX & Starbucks Corp. \\
Consumer Discretionary &  TGT & Target Corp. \\
Consumer Discretionary &  TJX  & TJX Cos Inc. \\
Consumer Discretionary &  TSCO  & Tractor Supply Company \\
Consumer Discretionary &  VFC  & VF Corp. \\
Consumer Discretionary &  WHR  & Whirlpool Corporation \\
Consumer Discretionary &  YUM  & Yum! Brands Inc. \\
Consumer Discretionary &  BWA  & BorgWarner Inc. \\
Consumer Discretionary &  CCL  & Carnival Corp. \\
Consumer Discretionary &  DRI &  Darden Restaurants Inc. \\
Consumer Discretionary &  DLTR  & Dollar Tree Inc. \\
Consumer Discretionary &  DHI & DR Horton, Inc. \\
Consumer Discretionary &  EBAY & eBay Inc. \\

\bottomrule
\end{tabular}

\begin{tabular}{lll}
\toprule
Sector &  Ticker & Name\\
\midrule
Consumer Staples       &  MO  & Altria Group Inc. \\
Consumer Staples       &  ADM  & Archer-Daniels-Midland Co. \\
Consumer Staples       &  EL & Estee Lauder Companies \\
Consumer Staples       &  GIS  & General Mills Inc. \\
Consumer Staples       &  HSY  & The Hershey Co. \\
Consumer Staples       &  HRL  & Hormel Foods Corp. \\
Consumer Staples       &  SJM  & JM Smucker Co. \\
Consumer Staples       &  K   & Kellogg Co. \\
Consumer Staples       &  KMB & Kimberly-Clark Corp. \\
Consumer Staples       &  KR    & The Kroger Co. \\
Consumer Staples       &  MKC      & McCormick \& Co. \\
Consumer Staples       &  TAP      & Molson Coors Beverage Co. \\
Consumer Staples       &  CPB     & Campbell Soup Co. \\
Consumer Staples       &  MNST    & Monster Beverage Corp. \\
Consumer Staples       &  PG    & Procter \& Gamble Co. \\
Consumer Staples       &  SYY  & Sysco Corp. \\
Consumer Staples       &  TSN   & Tyson Foods, Inc. \\
Consumer Staples       &  WMT    & Walmart Inc. \\
Consumer Staples       &  CHD    & Church \& Dwight \\
Consumer Staples       &  CLX    & Clorox Co. \\
Consumer Staples       &  KO      & Coca-Cola Co. \\
Consumer Staples       &  CL    & Colgate-Palmolive Company \\
Consumer Staples       &  CAG   & Conagra Brands Inc. \\
Consumer Staples       &  STZ   & Constellation Brands Inc. \\
Consumer Staples       &  COST  & Costco Wholesale Corp. \\
Energy                 &  APA   & APA Corp. \\
Energy                 &  BKR    & Baker Hughes \& Co. \\
Energy                 &  MRO    & Marathon Oil Corp. \\
Energy                 &  NOV    & Nov Inc. \\
Energy                 &  OXY    & Occidental Petroleum Corp. \\
Energy                 &  OKE    & ONEOK Inc. \\
Energy                 &  PXD     & Pioneer Natural Resources Co. \\
Energy                 &  SLB   & Schlumberger Nv \\
Energy                 &  VLO   & Valero Energy Corp. \\
Energy                 &  WMB     & The Williams Companies \\
Energy                 &  COG    & Coterra Energy Inc. \\
Energy                 &  CVX     & Chevron Corporation \\
Energy                 &  COP    & ConocoPhillips \\
Energy                 &  EOG   & EOG Resources Inc. \\
Energy                 &  XOM   & Exxon Mobil Corp. \\
Energy                 &  HAL   & Halliburton Co. \\
Energy                 &  HP     & HP Inc. \\
Energy                 &  HES    & Hess Corp. \\
\bottomrule
\end{tabular}

\begin{tabular}{lll}
\toprule
Sector &   Ticker & Name \\
\midrule
Financials             &  AFL     & Aflac Inc. \\
Financials             &  ALL     & Allstate Corp \\
Financials             &  SCHW    & Charles Schwab Corp. \\
Financials             &  CB       & Chubb Ltd. \\
Financials             &  CINF    & Cincinnati Financial Corp. \\
Financials             &  C       & Citigroup Inc. \\
Financials             &  CMA      & Comerica Inc. \\
Financials             &  RE       & Everest Re. Group \\
Financials             &  FITB    & Fifth Third Bancorp \\
Financials             &  BEN      & Franklin Resources Inc. \\
Financials             &  GL       & Globe Life Inc. \\
Financials             &  GS      & Goldman Sachs Group, Inc. \\
Financials             &  AXP    & American Express Company \\
Financials             &  HIG     & Hartford Financial Services Group \\
Financials             &  HBAN    & Huntington Bancshares Inc. \\
Financials             &  IVZ       & Invesco Ltd. \\
Financials             &  JPM    & JP Morgan \\
Financials             &  KEY     & KeyCorp \\
Financials             &  LNC     & Lincoln National Corp \\
Financials             &  MTB       & M\&T Bank Corp. \\
Financials             &  MMC     & Marsh \& McLennan Cos. \\
Financials             &  MCO     & Moody's Corp. \\
Financials             &  AIG     & American International Group Inc. \\
Financials             &  MS    & Morgan Stanley  \\
Financials             &  NTRS  & Northern Trust Corp.  \\
Financials             &  PBCT     & People's United Financial Inc.  \\
Financials             &  PNC      & PNC Financial Services Group  \\
Financials             &  PGR    & The Progressive Corp.  \\
Financials             &  RJF    & Raymond James Financial, Inc.  \\
Financials             &  SPGI   & S\&P Global Inc.  \\
Financials             &  STT    & State Street Corp.  \\
Financials             &  SIVB   & SVB Financial Group  \\
Financials             &  TROW   & T. Rowe Price Group  \\
Financials             &  AON    & Aon Plc  \\
Financials             &  TRV    & Travelers Companies Inc.  \\
Financials             &  TFC    & Truist Financial Corp  \\
Financials             &  UNM   & Unum Group  \\
Financials             &  USB   & US Bancorp  \\
Financials             &  WFC   & Wells Fargo \& Company  \\
Financials             &  ZION  & Zions Bancorp  \\
Financials             &  AJG   & Arthur J. Gallagher \& Co.  \\
Financials             &  BAC   & Bank of America Corp.  \\
Financials             &  BK    & Bank of New York Mellon Corp.  \\
Financials             &  BLK   & Blackrock Inc.  \\
Financials             &  COF   & Capital One Financial Corp.  \\

\bottomrule
\end{tabular}

\begin{tabular}{lll}
\toprule
Sector &    Ticker & Name  \\
\midrule
Health Care            &  ABT   & Abbott Laboratories  \\
Health Care            &  ABMD  & Abiomed Inc.  \\
Health Care            &  CAH   & Cardinal Health, Inc.  \\
Health Care            &  CERN  & Cerner Corp.  \\
Health Care            &  CI   & Cigna Corp.  \\
Health Care            &  COO  & The Cooper Companies, Inc.  \\
Health Care            &  CVS   & CVS Health Corp  \\
Health Care            &  DHR    & Danaher Corp.  \\
Health Care            &  DVA    & DaVita Inc.  \\
Health Care            &  XRAY   & Dentsply Sirona Inc.  \\
Health Care            &  LLY    & Eli Lilly \& Co.  \\
Health Care            &  GILD   & Gilead Sciences Inc.  \\
Health Care            &  A      & Agilent Technologies Inc.  \\
Health Care            &  HSIC   & Henry Schein Inc.  \\
Health Care            &  HOLX    & Hologic Inc.  \\
Health Care            &  HUM   & Humana Inc.  \\
Health Care            &  IDXX  & IDEXX Laboratories  \\
Health Care            &  INCY   & Incyte Corp.  \\
Health Care            &  JNJ    & Johnson \& Johnson  \\
Health Care            &  LH      & Laboratory Corp of America Holdings  \\
Health Care            &  MCK    & McKesson Corporation  \\
Health Care            &  MDT    & Medtronic PLC  \\
Health Care            &  MRK    & Merck \& Co., Inc.  \\
Health Care            &  ABC    & AmerisourceBergen Corp.  \\
Health Care            &  MTD   & Mettler-Toledo International Inc.  \\
Health Care            &  PKI   & PerkinElmer, Inc.  \\
Health Care            &  PFE   & Pfizer Inc.  \\
Health Care            &  DGX    & Quest Diagnostics  \\
Health Care            &  REGN  & Regeneron Pharmaceutical Inc.  \\
Health Care            &  RMD    & ResMed Inc.  \\
Health Care            &  STE    & Steris PLC  \\
Health Care            &  SYK    & Stryker Corp.  \\
Health Care            &  TFX    & Teleflex Inc.  \\
Health Care            &  TMO    & Thermo Fisher Scientific Inc.  \\
Health Care            &  AMGN   & Amgen Inc.  \\
Health Care            &  UNH   & UnitedHealth Group Inc.  \\
Health Care            &  UHS    & Universal Health Services Inc.  \\
Health Care            &  VRTX  & Vertex Pharmaceuticals Inc. \\
Health Care            &  WAT      & Waters Corporation \\
Health Care            &  BAX      & Baxter International Inc. \\
Health Care            &  BDX     & Becton Dickinson \& Co. \\
Health Care            &  BIIB    & Biogen Inc. \\
Health Care            &  BSX     & Boston Scientific Corp. \\
Health Care            &  BMY      & Bristol-Myers Squibb Co. \\
\bottomrule
\end{tabular}

\begin{tabular}{lll}
\toprule
Sector &   Ticker & Name \\
\midrule
Industrials            &  MMM      & 3M Company \\
Industrials            &  ALK     & Alaska Air Group Inc. \\
Industrials            &  CHRW    & CH Robinson Worldwide, Inc. \\
Industrials            &  CTAS   & Cintas Corp. \\
Industrials            &  CPRT       & Copart Inc. \\
Industrials            &  CMI        & Cummins Inc. \\
Industrials            &  DE      & Deere \& Co. \\
Industrials            &  DOV     & Dover Corp. \\
Industrials            &  ETN     & Eaton Corp. PLC \\
Industrials            &  EMR     & Emerson Electric Co. \\
Industrials            &  EFX      & EquifaX Inc. \\
Industrials            &  EXPD    & Expeditors International of Washington \\
Industrials            &  FAST    & Fastenal Co. \\
Industrials            &  FDX     & FedEx Corp. \\
Industrials            &  FLS     & Flowserve Corp. \\
Industrials            &  GD      & General Dynamics Corp. \\
Industrials            &  AME     & AMETEK Inc. \\
Industrials            &  GE      & General Electric Co. \\
Industrials            &  HON     & Honeywell International Inc. \\
Industrials            &  IEX     & IDEX Corp. \\
Industrials            &  ITW      & Illinois Tool Works Inc. \\
Industrials            &  J      & Jacobs Engineering Group Inc. \\
Industrials            &  JBHT   & JB Hunt Transport Services, Inc. \\
Industrials            &  JCI     & Johnson Controls International plc \\
Industrials            &  KSU     & Kansas City Southern \\
Industrials            &  LHX     & L3Harris Technologies Inc. \\
Industrials            &  LMT     & Lockheed Martin Corp. \\
Industrials            &  AOS     & AO Smith Corp. \\
Industrials            &  MAS     & Masco Corp. \\
Industrials            &  NSC     & Norfolk Southern Corp. \\
Industrials            &  NOC    & Northrop Grumman Corp. \\
Industrials            &  ODFL    & Old Dominion Freight Line Inc. \\
Industrials            &  PCAR    & PACCAR Inc. \\
Industrials            &  PH       & Parker-Hannifin Corp. \\
Industrials            &  PNR     & Pentair PLC \\
Industrials            &  PWR      & Peter Warren Automotive Holdings \\
Industrials            &  RTX      & Raytheon Technologies Corp. \\
Industrials            &  RSG      & Republic Services Inc. \\
Industrials            &  BA       & The Boeing Company \\
Industrials            &  RHI      & Robert Half International Inc.  \\
Industrials            &  ROK      & Rockwell Automation Inc. \\
Industrials            &  ROL      & Rollins, Inc.  \\
Industrials            &  ROP     & Roper Technologies Inc. \\
Industrials            &  SNA     & Snap-on Incorporated \\
Industrials            &  LUV     & Southwest Airlines Co. \\
Industrials            &  SWK     & Stanley Black \& Decker Inc. \\
Industrials            &  TXT     & Textron Inc. \\
Industrials            &  TT       & Trane Technologies PLC \\
Industrials            &  UNP    & Union Pacific Corporation \\
Industrials            &  CAT      & Caterpillar Inc. \\
Industrials            &  UPS    & United Parcel Service, Inc. \\
Industrials            &  URI     & United Rentals, Inc. \\
Industrials            &  WM      & Waste Management Inc. \\
\bottomrule
\end{tabular}

\begin{tabular}{lll}
\toprule
Sector &   Ticker & Name \\
\midrule
Industrials            &  WAB     & Westinghouse Air Brake Technologies \\
Industrials            &  GWW     & WW Grainger Inc. \\
Information Technology &  ADBE      & Adobe Inc. \\
Information Technology &  AKAM      & Akamai Technologies, Inc. \\
Information Technology &  GLW       & Corning Inc. \\
Information Technology &  FFIV      & F5 Inc. \\
Information Technology &  FISV      & Fiserv Inc. \\
Information Technology &  IT        & Gartner Inc. \\
Information Technology &  HPQ       & HP Inc. \\
Information Technology &  INTC      & Intel Corp. \\
Information Technology &  IBM       & International Business Machines \\
Information Technology &  INTU     & Intuit Inc. \\
Information Technology &  JKHY     & Jack Henry \& Associates, Inc. \\
Information Technology &  KLAC     & KLA Corp. \\
Information Technology &  APH      & Amphenol Corp. \\
Information Technology &  LRCX     & Lam Research Corp. \\
Information Technology &  MXIM     & Maxim Integrated Products Inc. \\
Information Technology &  MCHP     & Microchip Technology Inc. \\
Information Technology &  MSFT     & Microsoft Corp. \\
Information Technology &  MSI      & Motorola Solutions, Inc. \\
Information Technology &  NTAP     & NetApp Inc. \\
Information Technology &  NLOK     & NortonLifeLock Inc. \\
Information Technology &  NVDA     & NVIDIA Corporation \\
Information Technology &  PAYX      & Paychex Inc. \\
Information Technology &  QCOM     & Qualcomm Inc. \\
Information Technology &  ANSS     & ANSYS, Inc. \\
Information Technology &  SWKS     & Skyworks Solutions, Inc. \\
Information Technology &  SNPS       & Synopsys Inc. \\
Information Technology &  VRSN     & VeriSign Inc. \\
Information Technology &  XRX       & Xerox Holdings Corp. \\
Information Technology &  XLNX      & Xilinx Inc. \\
Information Technology &  ZBRA      & Zebra Technologies Corp. \\
Information Technology &  AAPL      & Apple Inc. \\
Information Technology &  AMAT      & Applied Materials, Inc. \\
Information Technology &  ADSK      & Autodesk, Inc. \\
Information Technology &  CSCO     & Cisco Systems Inc. \\
Information Technology &  CTXS     & Citrix Systems Inc. \\
Information Technology &  CTSH     & Cognizant Technology Solutions \\
\bottomrule
\end{tabular}

\begin{tabular}{lll}
\toprule
Sector &   Ticker & Name \\
Materials              &  APD      & Air Products \& Chemicals Inc. \\
Materials              &  ALB      & Albemarle Corp. \\
Materials              &  IP       & International Paper Co. \\
Materials              &  LIN      & Linde PLC \\
Materials              &  MLM      & Martin Marietta Materials, Inc. \\
Materials              &  NEM     & Newmont Corp. \\
Materials              &  NUE      & Nucor Corp. \\
Materials              &  PPG      & PPG Industries Inc. \\
Materials              &  SEE      & Seeing Machines Ltd. \\
Materials              &  SHW      & The Sherwin-Williams Co. \\
Materials              &  VMC     & Vulcan Materials Company \\
Materials              &  AVY       & Avery Dennison Corp. \\
Materials              &  BLL     & Ball Corp. \\
Materials              &  DD      & DuPont de Nemours, Inc. \\
Materials              &  EMN     & Eastman Chemical Company \\
Materials              &  ECL     & Ecolab Inc. \\
Materials              &  FMC     & FMC Corporation \\
Materials              &  FCX     & Freeport-McMoRan Inc. \\
Materials              &  IFF     & International Flavors \& Fragrances \\
Real Estate            &  ARE     & Aecon Group Inc. \\
Real Estate            &  AMT     & American Tower Corp. \\
Real Estate            &  IRM     & Iron Mountain \\
Real Estate            &  KIM     & Kimco Realty Corp. \\
Real Estate            &  MAA & Mid-America Apartment Communities \\
Real Estate            &  PLD   & Prologis Inc. \\
Real Estate            &  PSA   & Public Storage \\
Real Estate            &  O     & Realty Income Corp. \\
Real Estate            &  SBAC  & SBA Communications Corp. \\
Real Estate            &  SPG   & Simon Property Group, Inc. \\
Real Estate            &  SLG      & SL Green Realty Corp. \\
Real Estate            &  UDR    & UDR Inc. \\
Real Estate            &  AVB    & AvalonBay Communities Inc. \\
Real Estate            &  VTR     & Ventas Inc. \\
Real Estate            &  VNO     & Vornado Realty Trust \\
Real Estate            &  WELL    & Welltower Inc. \\
Real Estate            &  WY     & Weyerhaeuser Company \\
Real Estate            &  BXP    & Boston Properties, Inc. \\
Real Estate            &  DRE    & Duke Realty Corp. \\
Real Estate            &  EQR   & Equity Residential \\
Real Estate            &  ESS     & Essex Property Trust, Inc. \\
Real Estate            &  FRT    & Federal Realty Investment Trust \\
Real Estate            &  PEAK   & Healthpeak Properties Inc. \\
Real Estate            &  HST     & Host Hotels \& Resorts, Inc. \\
\bottomrule
\end{tabular}

\begin{tabular}{lll}
\toprule
Sector &   Ticker & Name \\
Utilities              &  AES    & AES Corp. \\
Utilities              &  AEE    & Ameren Corp. \\
Utilities              &  EIX     & Edison International \\
Utilities              &  ETR    & Entergy Corp. \\
Utilities              &  EVRG   & Evergy Inc. \\
Utilities              &  ES     & Eversource Energy \\
Utilities              &  FE     & FirstEnergy Corp. \\
Utilities              &  NEE      & NextEra Energy Inc. \\
Utilities              &  NI       & NiSource Inc. \\
Utilities              &  PNW      & Pinnacle West Capital Corp. \\
Utilities              &  PPL    & PPL Corp. \\
Utilities              &  PEG     & Public Service Enterprise Group Inc. \\
Utilities              &  AEP     & American Electric Power Co. \\
Utilities              &  SRE     & Sirius Real Estate Ltd. \\
Utilities              &  SO     & The Southern Co. \\
Utilities              &  WEC      & WEC Energy Group Inc. \\
Utilities              &  ATO     & Atmos Energy Corp. \\
Utilities              &  CNP     & CenterPoint Energy Inc. \\
Utilities              &  CMS    & CMS Energy Corp \\
Utilities              &  ED     & Consolidated Edison Inc. \\
Utilities              &  D        & Dominion Energy Inc. \\
Utilities              &  DTE     & DTE Energy Co. \\
Utilities              &  DUK      & Duke Energy Corp. \\
\bottomrule
\end{tabular}

\section*{Data availability statement}
All equity data is obtained from Bloomberg (\url{https://www.bloomberg.com})

\bibliographystyle{_elsarticle-num-names}
\bibliography{__References.bib}
\biboptions{sort&compress}
\end{document}